# Generic Tracking Specifications Translation From Time Domain to Frequency Domain

Javier JOGLAR-ALCUBILLA[1]

[1]Avionics Department, Barajas Institute, Avda.América 119, 28042, Madrid, Spain

jjoglar@educa.madrid.org

## Abstract

In certain types of robust control techniques, it is common having to deal with control problems where the specifications, described in the time domain, need to be translated to the frequency domain. This usually happens in techniques, such as Quantitative Feedback Theory, where the control problem is developed in the frequency domain. Therefore, not only process plants and disturbances should be specified in this domain, but also the limits and restrictions initially imposed in time. The question is important if we consider that any deviation in the parameters transfer from one domain to another will decisively influence in the development of the problem and, above all, in the finally result expressed in temporal terms. The technique presented allows the translation of the lower frequency limit in generic tracking specifications from time domain to frequency domain accurately. It will use approaches based on $2^{nd}$ order systems or an envelope approach based on higher order systems.

## 1 Introduction

Quantitative Feedback Theory or QFT (Houpis and Rasmussen, 1999; Yaniv, 1999) is a control methodology in the frequency domain and, therefore, the specifications required in the time domain should be transferred as closely as possible to the frequency domain. This problem has been considered from the beginning of the theory, as in (Horowitz and Sidi, 1972), in the further development of the Krishnan and Cruickshanks technique described in (Krishnan and Cruickshanks, 1977), in the model-based technique (D'Azzo and Houpis, 1995) or in the procedure given in (Franchek and Herman, 1998). All of them require manual iterative processes, ending when the designer decides that the results obtained are acceptable.

Despite these contributions, rigorous specifications transfer from one domain to another is still objectively an unsolved problem taking into account the above techniques are not too much precise in the conversion. It is often said this problem has presented little difficulty in QFT methodology (Kerr, 2004).

QFT control technique uses iterative procedures: after synthesizing the controller, if results obtained from temporal analysis are unsuitable, although the frequency specifications are fulfilled, you must seek another controller on a new redesign process; if the temporal results remain being inadequate, frequency specifications are adjusted and the design process start again; at the end, you can get a solution which fulfills temporal specifications. However, if we get a precise conversion time-frequency method, the synthesis procedure of the controller and/or precompensator will not need so much iteration. Also, it will provide solutions with less overdesign, a matter of great importance for the feedback cost (Gil-Martínez and García_Sanz, 2003).

In this paper, we present an automatic procedure for generic tracking specifications translation from time domain to frequency domain. It generates an automatic approach to a transfer function (*TF*) in the form of a $2^{nd}$ order subcritical damping system with parameters, "$\omega_n$" (natural frequency) and "$\zeta$" (damping coefficient) (Joglar and Aranda, 2014). To get it, we are applying a successive approximations technique, using inverse interpolation to obtain rise times and settling times imposed, as part of the overall process of translation.

Specifically, we use the "*Ascending and Descending Differences of $5^{th}$ order Newton Method*", described in detail in (Joglar-Alcubilla, 2015), which presents practical advantages (Iyengar, 2008). The proposed technique will work with step inputs, providing step responses with $2^{nd}$ order subcritical damping.

Also, it is developed a manual procedure to translate temporary tracking specifications to frequency domain, using an envelope approach, based on higher order systems.

## 2 Specifications translation

The method for specifications translation from "*time domain (TD)*" to "*frequency domain (WD)*" presented here (Joglar-Alcubilla, 2015) is described by the following relationship,

$$TD_{original}(Mp, tr, ts, dev, \omega_i)$$
$$\downarrow \qquad (1)$$
$$WD(Num_{min}, Den_{min}, Num_{max}, Den_{max})$$

Being the *TD* parameters (*Mp*, overshoot, *tr*, rise time, *ts*, settlement time; *dev*, settlement channel or admissible tolerance) maximum values of the response for tracking to a step input. With them, the *WD* parameters are obtained based on 2$^{nd}$ order subcritical damping *TFs*, defining $T_{Lo}(j\omega)$, the lower limit *TF*, given by,

$$T_{Lo}(j\omega) = \frac{Num_{min}}{Den_{min}}(j\omega) \qquad (2)$$

and $T_{Hi}(j\omega)$, the upper limit *TF*,

$$T_{Hi}(j\omega) = \frac{Num_{max}}{Den_{max}}(j\omega) \qquad (3)$$

$T_{Hi}(j\omega)$ is determined from the parameter $\omega_i$, which describes the number of times the natural frequency of the upper limit is greater than $\omega_n$, natural frequency, previously calculated for the lower limit $T_{Lo}(j\omega)$.

The steps sequence of the technique for generic tracking specifications translation is as follows,

1) Determination of minimum damping coefficient $\zeta_{min}$. We consider the overshoot *Mp* defined for underdamped systems as in (Ogata, 1993), i.e.:

$$Mp = e^{-\frac{\zeta\pi}{\sqrt{1-\zeta^2}}} \qquad (4)$$

Then $\zeta_{min}$ is given by the following equation, expressed in terms of *Mp*,

$$\zeta_{min} = \left[\frac{\left(\ln Mp/\pi\right)^2}{1+\left(\ln Mp/\pi\right)^2}\right]^{1/2} \qquad (5)$$

2) Find out the relationship $\omega_n = f(\zeta, ti)$, being $ti = (tr|ts)$. 2$^{nd}$ order systems with same $\zeta$, but different $\omega_n$, are characterized by the same maximum overshoot *Mp*, so they have the same relative stability.

The time constant of a 2$^{nd}$ order system defined by *T*, i.e.,

$$T = \frac{1}{\zeta\omega_n} \qquad (6)$$

can be expressed for *tr* and *ts*, respectively, as:

$$tr(\omega_n) = \frac{1}{\zeta\omega_n}k_r \qquad (7)$$

$$ts(\omega_n) = \frac{1}{\zeta\omega_n}k_s \qquad (8)$$

Where parameters $k_r$ and $k_s$ depend on the tolerance band value imposed (Ogata, 1993). Furthermore, for the particular case with $\omega_n=1$, (7) and (8) are rewritten as,

$$tr(\omega_n = 1) = \frac{1}{\zeta}k_r \qquad (9)$$

$$ts(\omega_n = 1) = \frac{1}{\zeta}k_s \qquad (10)$$

If (7) is combined with (9) and (8) with (10), any pair of natural frequency values $\omega_{nr}$ and $\omega_{ns}$, associated to $t_r$ and $t_s$, respectively, can be defined as,

$$\frac{tr(\omega_n = 1)}{tr(\omega_{nr})} = \omega_{nr} \qquad (11)$$

$$\frac{ts(\omega_n = 1)}{ts(\omega_{ns})} = \omega_{ns} \qquad (12)$$

2a) Then, applying equation (11) with *tr* y $\zeta_{min}$, we achieve,

$$\frac{tr(\omega_n = 1, \zeta_{min})}{tr(\omega_{nr}, \zeta_{min})} = \omega_{nr}(tr, \zeta_{min}) \qquad (13)$$

Being $tr(\omega_{nr},\zeta_{min})$ an input parameter and $tr(\omega_n=1,\zeta_{min})$ calculated with any adequate technique of inverse interpolation, the natural frequency $\omega_{nr}(tr,\zeta_{min})$ for $t_r$ is obtained from equation (13).

2b) Similarly, applying equation (12) with *ts* y $\zeta_{min}$, we achieve,

$$\frac{ts(\omega_n = 1, \zeta_{min})}{ts(\omega_{ns}, \zeta_{min})} = \omega_{ns}(ts, \zeta_{min}) \qquad (14)$$

With the input parameter $ts(\omega_{ns},\zeta_{min})$ and $ts(\omega_n=1,\zeta_{min})$ calculated with an adequate technique of inverse interpolation, the natural frequency $\omega_{ns}(ts,\zeta_{min})$ for $t_s$ is obtained from equation (14).

To determine $tr(\omega_n=1,\zeta_{min})$ in 2a) and $ts(\omega_n=1,\zeta_{min})$ in 2b), we apply "differences Newton method" (Bonnans et al., 2006), using as the interpolation function the step sign $f_{step}(t)$, i.e.:

$$f_{step}(t) = 1 - \frac{e^{-\zeta\omega_n t}}{\sqrt{1-\zeta^2}} sen\left(\omega_n\sqrt{1-\zeta^2}\,t + \arccos\zeta\right) \qquad (15)$$

Where $f_{step}(t_r)$ and $f_{step}(t_s)$ are given.

Observe that we want to determine temporal parameters by inverse interpolation in (15) with $\omega_n=1$ and $\zeta=\zeta_{min}$ so, only there is one unknown parameter in the equation each time.

3) Choose for $\omega_n(\zeta_{min})$ the greater value selected between $\omega_{nr}(tr,\zeta_{min})$ and $\omega_{ns}(ts,\zeta_{min})$. This value is the most restrictive condition for the temporary joint requirements *(tr|ts)*.

4) Calculate $\omega_n(\zeta,tr,ts)$ using $1>\zeta>\zeta_{min}$. From now, it will be named as $(\omega_n,\zeta)$ pairs or double vector $wd(\omega_n,\zeta)$. We get this with steps 2a), 2b) and 3), varying $\zeta$ from $\zeta_{min}$ to *1*, both values excluded. The step interval between a $\zeta$ value and the next one will determine the accuracy of the final result. Observe that for a specific $\zeta$ value, if you increase $\omega_n$ value respect to the one given for the associated pair $(\omega_n,\zeta)$, rise and settlement times make smaller. So, allowed frequency response for upper temporal limits is given for the area over the curve $(\omega_n,\zeta)$. See Figure 1.

5) Define 2$^{nd}$ order curves $T(j\omega)$, such that,

$$T(s=j\omega) = \frac{\omega_n^2}{s^2 + 2\zeta\omega_n s + \omega_n^2} \quad (16)$$

Where, replacing the double vector $wd(\omega_n,\zeta)$, we obtain the set of curves in the Bode diagram, fulfilling tracking specifications for the upper time limit, initially given.

6) Define 2$^{nd}$ order curves $T'(j\omega)$, such that with $\omega_{ni}=i\omega_n$ and $i=1,2,..,\omega_i$

$$T'(s=j\omega) = \frac{\omega_{ni}^2}{s^2 + 2\zeta\omega_{ni} s + \omega_{ni}^2} \quad (17)$$

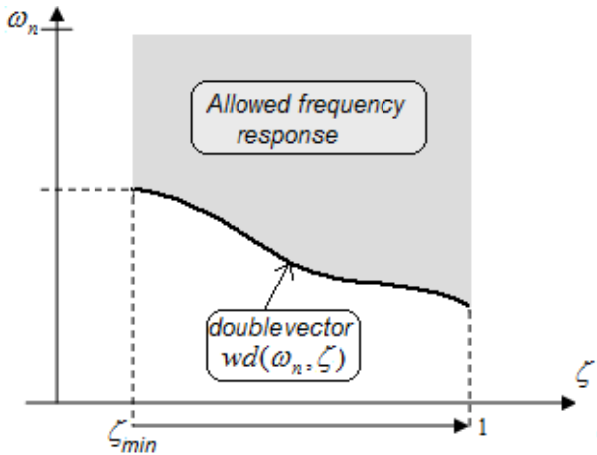

**Figure 1.** Graphical representation of $wd(\omega_n,\zeta)$

Thus, substituting $wd(\omega_{ni},\zeta)$ in (17), $\omega_i$ sets of curves in the Bode diagram are obtained, fulfilling both tracking requirements, the upper temporal limit $(\omega_{ni}=\omega_n)$ and the lower temporal limit $(\omega_{ni}=\omega_i\omega_n)$.

7) Achieving the lower and upper limits in the frequency domain, $T'_{Lo}(j\omega)$ and $T'_{Hi}(j\omega)$. On the Bode diagram above is sought, on the one hand, the lower intersection (minimums in magnitude and in phase) and, secondly, the upper intersection (maximums in magnitude and in phase) of the different curves defined by the vectors sets $wd(\omega_{ni},\zeta)$.

8) As the lower and upper frequency limits ($T'_{Lo}(j\omega)$ and $T'_{Hi}(j\omega)$) are composed of the intersection of several curves $T'(j\omega)$, the way to describe both *TFs* is to approximate them using one of the following criteria, depending on the bandwidth and accurate of interest. So, we obtain $T_{Lo}(j\omega)$ and $T_{Hi}(j\omega)$:
  a) Considering *restriction at low frequencies*.
  b) Considering *restriction at high frequencies*.
  I.e., choose the curves with minimum ($min|T'(j\omega)|$) and maximum ($max|T'(j\omega)|$) magnitude in the curves set $wd(\omega_{ni},\zeta)$, taking into account (a) the lowest or (b) the highest frequencies, respectively, into the work frequency band used.
  c) Considering *restriction at the envelope approach*. For accurate, we can seek *TFs* of the lower and upper limits $T_{Lo}(j\omega)$ and $T_{Hi}(j\omega)$. The results obtained in 7) are approximated by higher order models using an order reduction method, applied as in (Aranda and Joglar, 2014; Joglar-Alcubilla and Aranda, 2014).

9) Getting the final *TD* parameters, from the *WD* ones achieved, which match to the original *TD* requirements. That is,

$$TD_{original}(Mp_{max},tr_{max},ts_{max},dev,\omega_i)$$
$$\downarrow$$
$$WD(Num_{min},Den_{min},Num_{max},Den_{max}) \quad (18)$$
$$\downarrow$$
$$TD_{final}(Mp_{upp},tr_{upp},ts_{upp},Mp_{low},tr_{low},ts_{low})$$

## 3 Envelope approach

We have developed a manual procedure to translate generic temporary tracking specifications to frequency domain, using an envelope approach based on reduction of higher order systems.
The method consists of rationalizing the envelopes $T'_{Lo}(j\omega)$ and $T'_{Hi}(j\omega)$, which represent the limits in frequency of the tracking specifications. These envelopes are expressed in complex form, within a certain frequency range, and must be approximated to *TFs* with not too high order: the reduced *TFs* solution will be $T_{Lo}(j\omega)$ and $T_{Hi}(j\omega)$.
Suppose one envelope $T'(j\omega)$ ($T'_{Lo}(j\omega)$ or $T'_{Hi}(j\omega)$) to rationalize. For this, consider the relationship between the envelope input in complex form $T'(j\omega)$ and its approach, the reduced transfer function $T(j\omega)$:

$$T(s) = \frac{b_0+b_1 s+...+b_n s^n}{a_0+a_1 s+...+a_m s^m} \approx T'(s) \text{, with } s=jw \quad (19)$$

Where the different parameters are defined as,
- *T'*: Complex input *TF* with [N,1] vector size.
- *w*: Range of frequencies with [N,1] vector size.

- *m* and *n*: Number of poles and zeros, respectively, for the reduced *TF* expressed in the form "numerator/denominator".

The equation (19) can be rewritten as:

$$[a_0 + a_1 s + \ldots + a_{m-1} s^{m-1}] - \frac{1}{T'}[b_0 + b_1 s + \ldots + b_n s^n] = -a_m s^m \quad (20)$$

From the input function *T'(s)* with N complex numbers, one for each frequency *w*, we obtain N equations, by replacing in (20). Applying the rationalization process described by (Horowitz, 1992), we obtain values for the *m* coefficients $a_i$ and *n* coefficients $b_i$.

Finally, we must use a polynomial evaluation over *T(s)*. This evaluation is used to compare in magnitude and phase with the original complex *TF T'(s)* value. So, the differences obtained are the errors in magnitude and phase of the process.

The practical development of this rationalization process, as it is described above, is the program RACWE.M. This can be downloaded as it is indicated in section 5.

The RACWE.M program is technically described as follows, in Matlab format:

$$\text{function [numer, denom]}= racwe (T', w, n, m) \quad (21)$$

We get working with complex input *T'* to offer as output the corresponding transfer function (numer/denom), according to the supplied frequency vector *w*. It also allows users to decide the order of the output functions by adjusting the number of poles and zeros they may contain. Depending on the order of the output *TF* we have selected, so it will be the difference in magnitude and phase for each frequency within the range used, between the output and the input. Therefore, selecting different output orders and observing the differences in magnitude and phase (errors), we can achieve reducing the order of the input function to a value where these differences are not too high and the approach adequate.

On the other hand, we must remind that the number of poles selected must be equal or greater than the number of zeros, to assure stability.

Once the output *TF* is obtained the user must decide, if it is necessary, to apply gain adjustment.

To maintain stability there must be no RHP poles or zeros in the final output *TF*, nor poles with zero value. So, the program allows eliminating this type of poles/zeros or even those with insignificant values.

The program offers numerical and graphical magnitude/phases differences between the input *T'(jω)* and the output *T(jω)*. With the maximum magnitude and phase errors the user decides if the approach is adequate or not.

The process ends offering numerical and graphical temporary responses associated with *T(jω)* obtained.

# 4 Example application

Suppose tracking specifications given by the lower limit and the upper limit, described by the set of temporal parameters $TD_{original}(15\%,5s,30s,\pm3\%,5)$, i.e., 15% of maximum overshoot, maximum rise and settlement times of 5s and 30s, respectively, settlement channel deviation ±3% and upper frequency limit defined as $5\omega_n$. Applying the technique described above, and using the programs developed in Matlab indicated below in section 5, we obtain:

- 2$^{nd}$ order responses *T(jω)*, from $wd(\omega_n, \zeta)$, which are fitted to the $TD_{original}$ parameters, given by the upper time limit; these are described by the following *TFs*,

$$\frac{0.1137}{s^2 + 0.3486s + 0.1137}, \frac{0.1284}{s^2 + 0.4063s + 0.1284}, \frac{0.1459}{s^2 + 0.4713s + 0.1459},$$

$$\frac{0.1668}{s^2 + 0.5448s + 0.1668}, \frac{0.1918}{s^2 + 0.6279s + 0.1918}, \frac{0.2211}{s^2 + 0.7212s + 0.2211}, \frac{0.21}{s^2 + 0.7487s + 0.21}$$

$$\frac{0.3013}{s^2 + 0.9518s + 0.3013}, \frac{0.3923}{s^2 + 1.149s + 0.3923}, \frac{0.5438}{s^2 + 1.426s + 0.5438} \quad (19)$$

- Sets of 2$^{nd}$ order responses *T'(jω)*, from $wd(\omega_{ni}, \zeta)$. Results in Figure 2.
- Sets of *TD* responses, equivalent to those *T'(jω)* systems. Results in Figure 3.
- Lower and upper limits in the frequency domain, *T'$_{Lo}$(jω)* and *T'$_{Hi}$(jω)*. These are obtained looking for the minimum and maximum magnitudes/phases, respectively, for each frequency. Results in Figure 4.
- Lower and upper limits in the frequency domain, *T$_{Lo}$(jω)* and *T$_{Hi}$(jω)*, considering *restriction at low frequencies*, i.e., from the curves of minimum and maximum magnitude, respectively, in the low frequency area. Results in Figure 5.

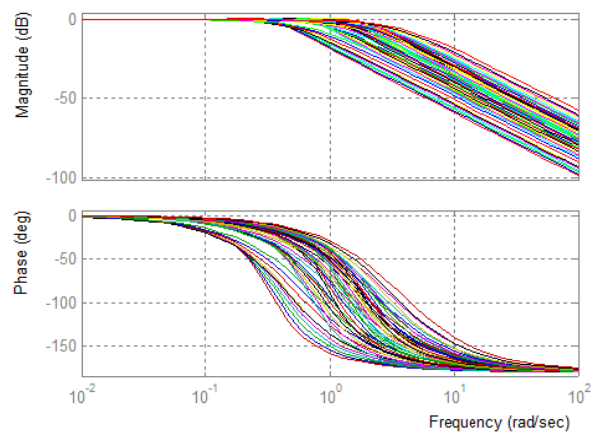

**Figure 2.** 2$^{nd}$ order Responses *T'(jω)*

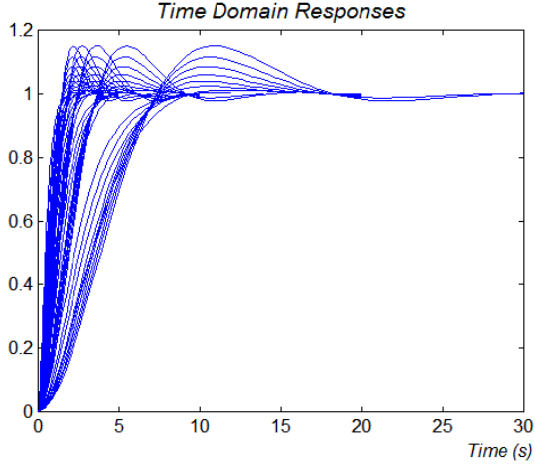

**Figure 3.** Set of temporary Responses, equivalent to the systems $T'(j\omega)$

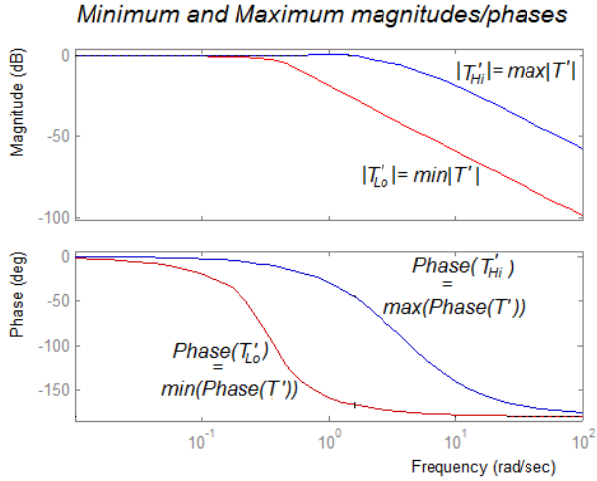

**Figure 4.** Lower and upper limits in the frequency domain (Bode diagram), $T'_{Lo}(j\omega)$ and $T'_{Hi}(j\omega)$

- Curves and *final TD* parameters which fit to the *original TD*, applying restriction at low frequencies. Temporary results in Figure 6.
  - ✓ Lower frequency limit defined as the initial tightening curve $T(j\omega)$, at low frequencies,

$$\left(\frac{Num_{min}}{Den_{min}}\right)_{Lo} = \frac{0.3923}{s^2 + 1.149s + 0.3923} \quad (20)$$

  - ✓ Upper frequency limit defined as the most restricted $T(j\omega)*5\omega_n$ curve, at low frequencies,

$$\left(\frac{Num_{max}}{Den_{max}}\right)_{Lo} = \frac{2.843}{s^2 + 1.743s + 2.843} \quad (21)$$

- Lower and upper limits in the frequency domain, $T_{Lo}(j\omega)$ and $T_{Hi}(j\omega)$, considering *restriction at high frequencies*, i.e., from the curves of minimum and maximum magnitude, respectively, in the high frequency area. Results in Figure 7

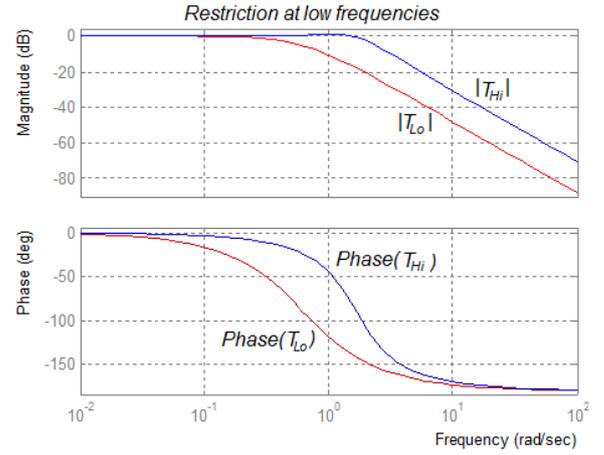

**Figure 5.** Lower and upper limits in the WD, $T_{Lo}(j\omega)$ and $T_{Hi}(j\omega)$, considering restriction at low frequencies.

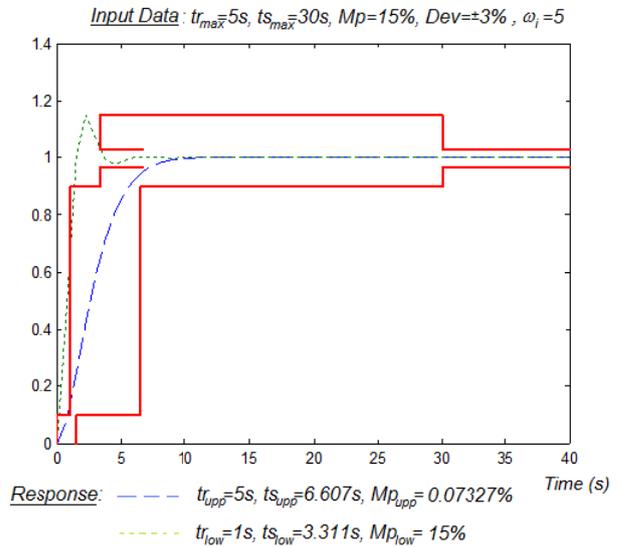

**Figure 6.** Final temporary responses associated with $T_{Lo}(j\omega)$ and $T_{Hi}(j\omega)$, considering restriction at low frequencies.

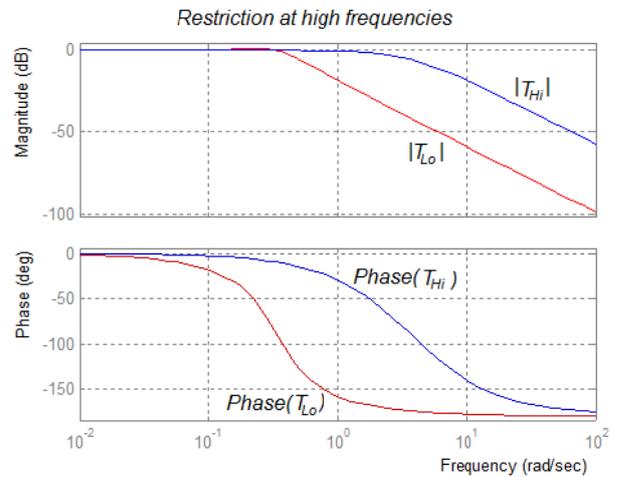

**Figure 7.** Lower and upper limits in the WD, $T_{Lo}(j\omega)$ and $T_{Hi}(j\omega)$, considering restriction at high frequencies.

- Curves and *final TD* parameters which fit to the *original TD,* applying restriction at high frequencies. Temporary results in Figure 8.
  - Lower frequency limit defined as the initial tightening curve *T(jω)* , at high frequencies,

  $$\left(\frac{Num_{min}}{Den_{min}}\right)_{Hi} = \frac{0.1137}{s^2 + 0.3486s + 0.1137} \quad (22)$$

  - Upper frequency limit defined as the most restricted *T(jω)*5ω$_n$ curve, at high frequencies,

  $$\left(\frac{Num_{max}}{Den_{max}}\right)_{Hi} = \frac{13.59}{s^2 + 7.13s + 13.59} \quad (23)$$

- Lower and upper limits in the frequency domain, *T$_{Lo}$(jω)* and *T$_{Hi}$(jω)*, considering *restriction at the envelope approach*, i.e., rationalizing real limits *T'$_{Lo}$(jω)* and *T'$_{Hi}$(jω)* until *TFs* have not got too high order. Results in Figure 9.

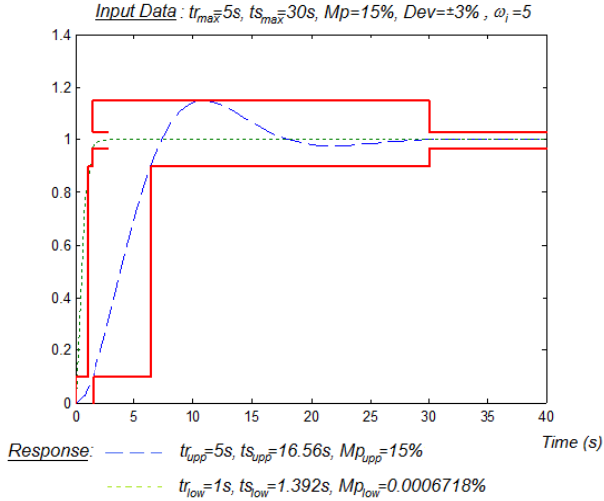

**Figure 8.** Final temporary responses associated with *T$_{Lo}$(jω)* and *T$_{Hi}$(jω)*, considering restriction at high frequencies.

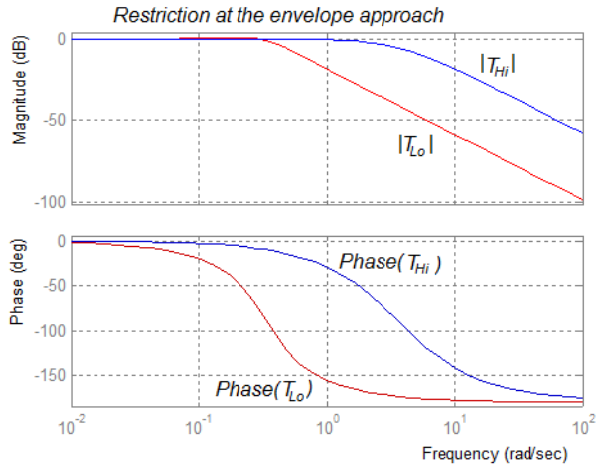

**Figure 9.** Lower and upper limits in the *WD*, *T$_{Lo}$(jω)* and *T$_{Hi}$(jω)* , considering restriction at the envelope approach.

- Curves and *final TD* parameters which fit to the *original TD,* applying restriction at the envelope approach. Temporary results in Figure 10.
  - Lower frequency limit *T$_{Lo}$(jω)* defined by rationalizing the lower intersection in the set *T'(jω)*.

  $$\left(\frac{Num_{min}}{Den_{min}}\right)_{Env} = \frac{0.1168}{s^2 + 0.3903s + 0.1168} \quad (24)$$

  Graphical magnitude/phases differences between *T'$_{Lo}$(jω)* and *T$_{Lo}$(jω)* are expressed in Figure 11. We are obtaining a maximum magnitude error of *0.131rad* and a maximum phase error of *-3.6015 degrees*.
  - Upper frequency limit *T$_{Hi}$(jω)* defined by rationalizing the lower intersection in the set *T'(jω)*.

  $$\left(\frac{Num_{max}}{Den_{max}}\right)_{Env} = \frac{0.002s + 13}{s^2 + 6.79s + 13} \quad (25)$$

  Graphical magnitude/phases differences between *T'$_{Hi}$(jω)* and *T$_{Hi}$(jω)* are expressed in Figure 12. We are obtaining a maximum magnitude error of *0.1799rad* and a maximum phase error of *1.7387 degrees*.

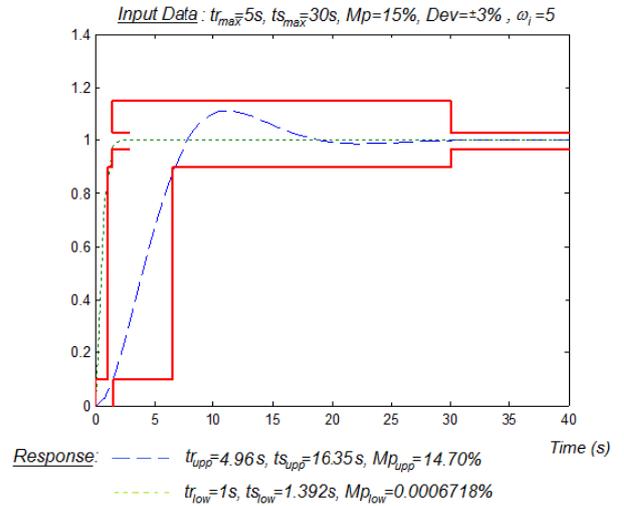

**Figure 10.** Final temporary responses associated with *T$_{Lo}$(jω)* and *T$_{Hi}$(jω)*, considering restriction at the envelope approach.

# 5 Application development

The author has developed in Matlab the necessary programs to verify the above example application and any other, for tracking temporary specifications translation to frequency domain.
"Tracking specifications translation from time domain to frequency domain" can be downloaded

from the next URL:
https://www.dropbox.com/sh/x7ywzymrkkr2qks/AABC2dvUDAfVPWBubWzo-qWQa?dl=0

The application starts with the program TD2WD.M

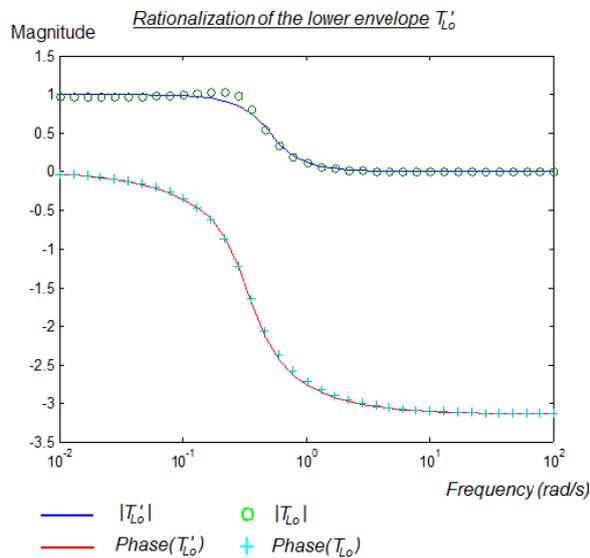

**Figure 11.** Graphical differences in magnitude and phase between $T'_{Lo}(j\omega)$ and the lower envelope approach TF $T_{Lo}(j\omega)$.

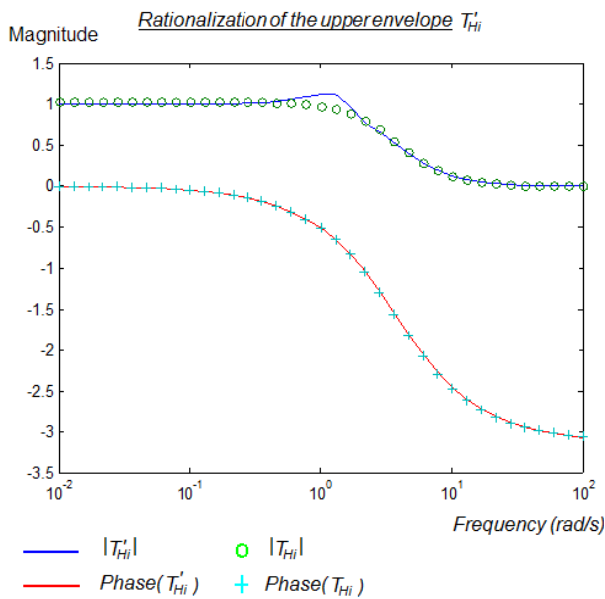

**Figure 12.** Graphical differences in magnitude and phase between $T'_{Hi}(j\omega)$ and the upper envelope approach TF $T_{Hi}(j\omega)$.

# 6 Conclusions

The contributions made for the purpose of reliable and accurate conversion of temporary specifications to the frequency domain in different fields of control, as in QFT, have been insufficient, so that, rigorous specifications translation from time to frequency remains being actually an unsolved problem.

In practice, it has not paid much attention to this problem: control methodologies affected, including QFT, often use iterative processes for synthesizing the controller, ending in a temporary analysis; if the temporary verification is not adequate, although frequency specifications are fulfilled, the designer goes back in a new redesign process of the controller; ultimately, to get good temporary results, frequency specifications are relaxed, starting the process again.

The technique presented here allows the conversion of generic tracking specifications from time to frequency accurately, so that, the process of synthesizing the controller in the design phase will not require much iteration; also, it will be achieved solutions with less overdesign: consider, for example, that in the methodology QFT, the precision degree of specifications translation influences in the level of accuracy of the bounds, at least, in the final design stage.

It uses variation of the maximum temporary input parameters, overshoot, rise time and settling time, by inverse interpolation, to find the most appropriate frequency response to the type of selected approach: maximum accuracy in a specific area of the bandwidth (low or high frequency with $2^{nd}$ order systems) or in the entire bandwidth (envelope with reduced higher order systems). Once the lower frequency limit (temporary response with maximum values) is defined, the upper frequency limit is determined, proportional to previous one in the *WD*.

The upper frequency limit associated with tracking specifications is often described in temporary terms of maximum overshoot response to a given step input. From the technique presented here, it can be developed a specific method to get this kind of tracking specification in the *WD*.

Instead of using inverse interpolation methods of temporary parameters, relatively complex, the methodology developed can be simplified applying classical hypothesis, but at the cost of reducing its accuracy significantly.

Furthermore, the methodology for translation of temporary specifications to frequency ones using inverse interpolation can be applied also in those situations where a step or pulse input must produce a damped impulse response. This case corresponds to the specifications of sensitivity or decoupled tracking.

# References

C. Houpis, S.J. Rasmussen. Quantitative Feedback Theory: fundamentals and applications. New York, USA: *Marcel Dekker*. 1999.

O. Yaniv. Quantitative feedback design of linear and nonlinear control systems. *Kluwer Academic Publishers*, 1999, Norwell, Massachusetts.

I.M. Horowitz, M. Sidi. Synthesis of feedback systems with large plant ignorance for prescribed time domain tolerances. *International Journal of Control*, 1972, vol. 16, pp. 287-309.

K.R. Krishnan, A. Cruickshanks. Frequency-domain design of feedback systems for specified insensitivity of time-domain response to parameter variation. *International Journal of Control*, 1977, vol. 25, no 4, pp. 609-620, 1977.

J. D'Azzo, C. Houpis. Linear control system analysis and design conventional and modern. New-York, USA: *McGraw-Hill*, 1995, 4ªEd, Cap18.

M.A. Franchek, P.A. Herman. Direct connection between time-domain performance and frequency-domain characteristics. *International Journal of Robust and Nonlinear Control*, 1998, vol. 8, pp. 1021-1042.

M. Kerr. Robust control of an articulating flexible structure using MIMO QFT. Thesis, 2004, University of Queensland.

M. Gil-Martínez, M. García_Sanz. Simultaneous meeting of robust control specifications in QFT. *International Journal of Robust and Nonlinear Control*, 2003, 13(7), pp. 643-656.

J. Joglar, J. Aranda. QFT multivariable control for the longitudinal dynamics of an air vehicle. *The 9th lnternational Conference on Electrical and Control Technologies*, 2014, ECT2014, pp.26-29, Lithuania.

J. Joglar-Alcubilla. Contribución al estudio de diseño de sistemas de control mediante QFT: aplicaciones al diseño de sistemas de control de vuelo y navegación. *Thesis*.2015, Madrid, Spain: ETSI Informática, UNED.

S.R.K. Iyengar. Numerical methods and computation. Interpolation and approximation. Lecture Nº 28. Dep. of Mathematics, Indian Institute of Technology, Delhi. 2008.

K. Ogata. Ingeniería de control moderna. Prentice Hall. 2ºedition, 1993.

J.F. Bonnans, J.C. Gilbert, C. Lemaréchal, C.A. Sagastizábal. Numerical optimization: Theoretical and practical aspects. Universitext (Second revised ed. of translation of 1997 French ed.). Berlin: *Springer-Verlag*. pp. xiv+490. DOI: 10.1007/978-3-540-35447-5. ISBN 3-540-35445-X. MR 2265882. 2006.

J. Aranda, J. Joglar. Application of QFT order reduction methodology in the control of an autonomous marine vehicle. *Journal of Maritime Research,* 2014, Vol 11(3), pp. 67-78.

J. Joglar-Alcubilla, J. Aranda. Order reduction of transfer functions in practical control systems. *The 9th lnternational Conference on Electrical and Control Technologies*, 2014, ECT2014, pp. 30-35, Lithuania.

I.M. Horowitz. The Quantitative Feedback Theory, vol.1. QFT Publishers, Boulder, Colorado, 1992.